\documentclass{article}
\usepackage[preprint]{spconf}
\copyrightnotice{\copyright\ IEEE 2023}
\usepackage{amsmath, graphicx, booktabs, hyperref}
\usepackage [english]{babel}
\usepackage [autostyle, english = american]{csquotes}

\MakeOuterQuote{"}

\hypersetup{
    colorlinks=true,
    linkcolor=blue,    
    urlcolor=blue,
    pdfpagemode=FullScreen,
    }
\title{From sound to meaning in the auditory cortex: A neuronal representation and classification analysis}

\name{Kumar Neelabh, Vishnu Sreekumar\thanks{This work was supported in part by the Faculty Seed Grant provided by IIIT Hyderabad to VS. Icons by Freepik via Flaticon.}}
\address{Cognitive Science Lab, International Institute of Information Technology, Hyderabad, India.}

\begin{document}

\noindent\fbox{%
    \parbox{\textwidth}{%
        This work has been submitted to the IEEE for possible publication. Copyright may be transferred without notice, after which this version may no longer be accessible.
    }%
}

\maketitle

\begin{abstract}
The neural mechanisms underlying the comprehension of meaningful sounds are yet to be fully understood. While previous research has shown that the auditory cortex can classify auditory stimuli into distinct semantic categories, the specific contributions of the primary (A1) and the secondary auditory cortex (A2) to this process are not well understood. We used songbirds as a model species, and analyzed their neural responses as they listened to their entire vocal repertoire (\(\sim \)10 types of vocalizations). We first demonstrate that the distances between the call types in the neural representation spaces of A1 and A2 are correlated with their respective distances in the acoustic feature space. Then, we show that while the neural activity in both A1 and A2 is equally informative of the acoustic category of the vocalizations, A2 is significantly more informative of the semantic category of those vocalizations. Additionally, we show that the semantic categories are more separated in A2. These findings suggest that as the incoming signal moves downstream within the auditory cortex, its acoustic information is preserved, whereas its semantic information is enhanced.
\end{abstract}
\section{Introduction}
\label{sec:intro}

For vocal communication to be successful, it is essential for the brain to convert auditory input into meaningful percepts making it possible to decode meaning from neural firing patterns\cite{Robotka_Thomas_Yu_Wood_Elie_Gahr_Theunissen_2023, Elie_Theunissen_2015}. Such neural semantic categorization is a complex task because it requires neural representations that can identify invariant features across different speakers and renditions \cite{Elie_Theunissen_2019}. Previous work has shown that the auditory cortex can categorize vocal stimuli into distinct groups based on the meaning of those vocalizations \cite{Robotka_Thomas_Yu_Wood_Elie_Gahr_Theunissen_2023}. However, the specific computational contributions of the two sub-regions, the primary (A1) and the secondary auditory cortex (A2), are poorly understood \cite{Robotka_Thomas_Yu_Wood_Elie_Gahr_Theunissen_2023}.

\begin{figure}[t]
  \centering
  \includegraphics[width=\linewidth, trim=4 4 4 4, clip]{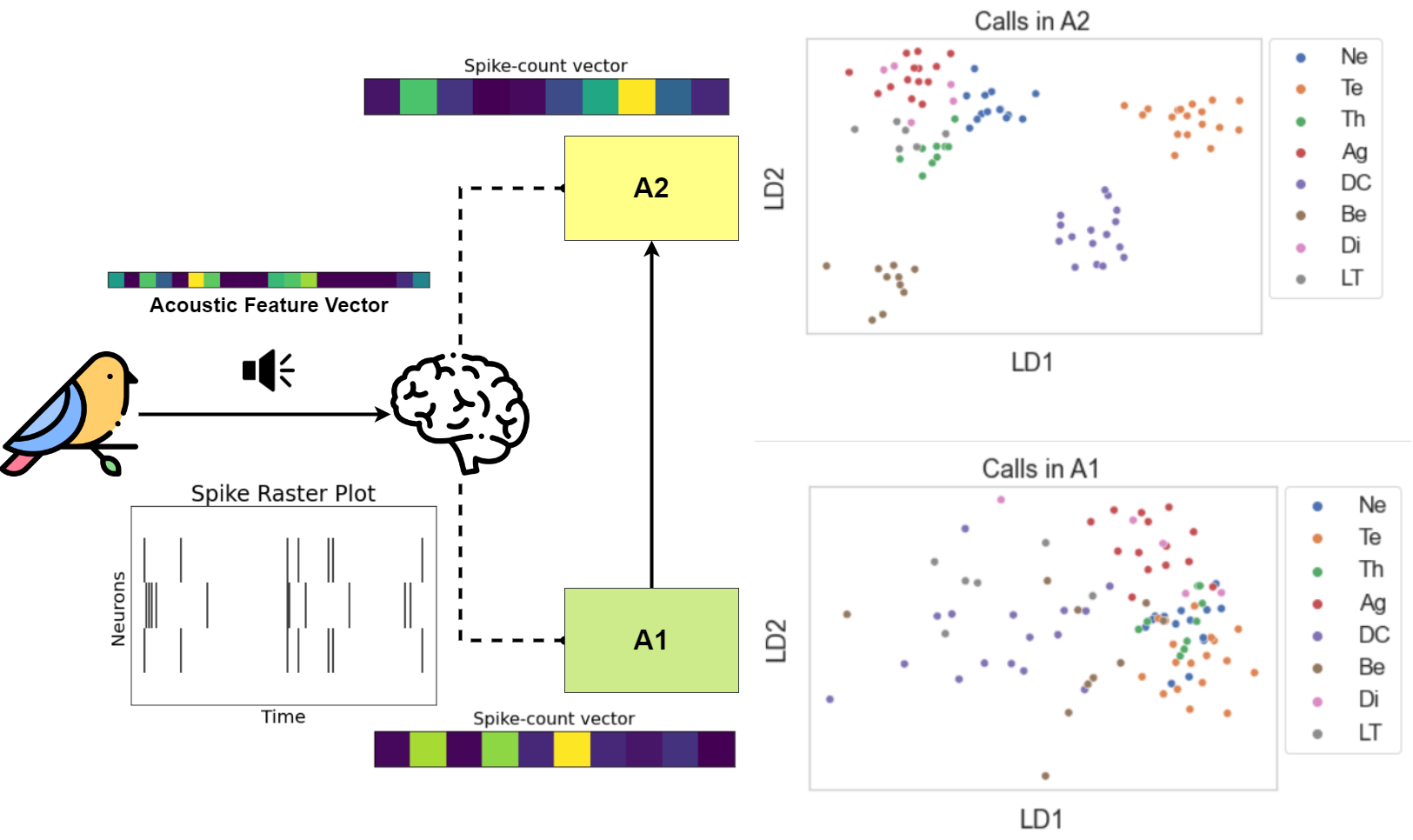}
  \caption{Overview: Calls incoming to the cortex land on A1 and are then passed to A2. Calls can be represented as acoustic feature vectors. The spike-raster plot shows timings of spikes fired by 3 neurons in response to a call. Spike count vectors are formed by summing the spikes for each neuron. The sub-figures to the right represent LDA projections of neural responses to calls. Each marker represents one audio file containing one call recording. Markers are labeled by the semantic category of the calls. Notice that the semantic categories are more separated in A2.}
  \label{fig:fig-1}
\end{figure}

Zebra finches, a species of songbird, are used as a model organism in the study of the neural basis of vocal communication because they possess brain regions analogous to A1 and A2 in humans \cite{Meliza_Margoliash_2012}. These birds are known to use 10 distinct types of vocalizations, each with a specific behavioral meaning, such as begging for food or alerting others to danger \cite{Elie_Theunissen_2016}. Because zebra finches have a relatively simple and well-defined vocal communication system, they provide a valuable model for investigating the neural mechanisms of auditory perception in a controlled and accessible manner.

We hypothesize that A2 (which lies downstream of A1) contains more information about the perceived meaning of vocalizations compared to A1 (Fig. \ref{fig:fig-1}). This is based on previous findings which indicate that increasingly more abstract categories of information are encoded as one moves along the ventral auditory pathway from A1 to A2 to the pre-frontal cortex (pFC) \cite{Tsunada_Cohen_2014}. Robotka et al (2023) \cite{Robotka_Thomas_Yu_Wood_Elie_Gahr_Theunissen_2023} reported no significant difference in semantic classification accuracy using either A1 or A2 activity in anesthetized birds. In awake birds, they report a higher accuracy using A1 activity. Their analyses are based on considerably smaller neural ensembles of 20 neurons. Since this result is unexpected based on what we already know about A1 and A2, we hypothesized that these findings may be a result of small ensemble size.

\section{Methods}
\label{sec:methods}

\subsection{Dataset and Pre-processing}
\label{subsec:preproc}
An open dataset of extracellular electrophysiological recordings of anesthetized zebra finches (N=5) listening to conspecific (i.e., belonging to the same species) vocalizations was used for this study \cite{Elie_Theunissen_Dataset}. Details about the dataset can be found in the original paper \cite{Elie_Theunissen_Dataset} but briefly, each bird was presented with $114 \pm 22$ (mean $\pm$ SEM) stimuli. Each stimulus was repeated $10$ times. Spike-sorting yielded $169.40 \pm 84.36$ (mean $\pm$ SEM) single units for each bird. Discrete spikes were smoothed using a 50 ms boxcar window to obtain firing rates. Time-varying population response of a given area (A1/A2) to a stimulus was represented as a matrix, with the trial-averaged firing rates of neurons in the area stacked in columns. Static population response was represented as a vector of trial-averaged spike counts.

In all the comparative analyses, equal-sized neural populations from A1 and A2 in the same hemisphere in the same bird were compared. E.g., to compare classification accuracy in A1 and A2 for populations of unequal sizes (say, 20 neurons in A1 but 60 neurons in A2), we take 100 random sample populations of size 20 from A2, and then average their classification accuracy. This gives us an estimate of the classification performance of a similarly-sized population in A2. Data and code are available at \url{https://github.com/mandalab/BirdSongNeuralRep}.

\subsection{Distance matrix comparison}
\label{methods:distance-matrices}
Here, we aimed to determine whether neuronal spiking patterns in A1 and A2 represent acoustic features. To do this, we tested whether the distances between semantic categories in the acoustic feature space were correlated with the corresponding distances in the neural representation spaces of A1 and A2. For a neural population that only encodes acoustic features, we can expect the two distance matrices to be correlated \cite{Kriegeskorte_Mur_Bandettini_2008}.  Here, we represented stimulus audio files as feature vectors. We used a set of 20 features that fall broadly into 3 categories: pitch features, amplitude features, and spectral features \cite{Elie_Theunissen_2016}. To enable a comparison between acoustic feature vectors and neuronal firing, we represented the corresponding neural responses in A1 and A2 as spike-count vectors. 

Distances between any given pair of vectors in a given space (acoustic/A1/A2) was given by the Euclidean metric. The distance between a pair of semantic categories in a given space was given as the Hausdorff distance between the two corresponding sets of vectors. Hausdorff distance between any two sets is given by the maximum of minimum distances from points in one set to those in the other set (Eq. \ref{eq:hausdorff}). Hausdorff distance is a commonly used method to find distances between two sets of points in high dimensional spaces \cite{Taha_Hanbury_2015}. In contrast to the average of pairwise distances, the Hausdorff distance is sensitive to positions of individual points in the two sets \cite{Taha_Hanbury_2015}. In addition, the Hausdorff has the useful mathematical property that the distance from a set to itself is 0 (also not satisfied by the average of pairwise distances). Finally, the correlation between distance matrices was evaluated using the Mantel Test.

\begin{equation} \label{eq:hausdorff}
H(A, B) = \mathop{max}_{a \in A} \mathop{min}_{b \in B} ||a - b||
\end{equation}

\begin{figure}[ht]
  \centering
  \includegraphics[width=\linewidth]{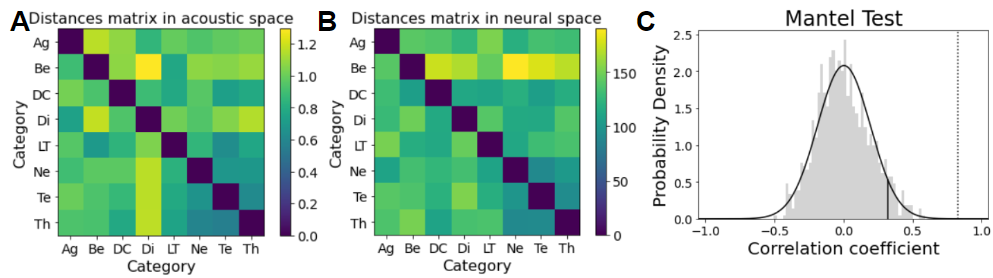}
  \caption{Results of the method validation experiment. Distances between semantic categories in the (A) acoustic feature space and (B) synthetic neural space are highly correlated. (C) Histogram of correlation coefficients obtained in Mantel test permutations. Dotted line indicates observed value.}
  \label{fig:validation-results}
\end{figure}

We created synthetic spike trains that were designed to be a good representation of the acoustic features in order to validate the distance matrix comparison method and show that the distance matrices obtained are indeed correlated.  We took a sample of 24 audio clips from the dataset and created a population of synthetic neurons such that each synthetic neuron responded to a particular acoustic feature of an audio clip. Specifically, the spike train emitted by a given neuron in response to a given stimulus was generated based on a Poisson distribution that describes its probability of firing at any given timestep (Eq. \ref{eq:poisson}). The Poisson distribution itself was parametrized by the value of the corresponding acoustic feature, by setting the \textit{\textmu} parameter equal to the value of the acoustic feature (with 50\% Gaussian noise). Note that the acoustic features were scaled so that the resulting spiking rates match those observed in the dataset. We applied our distance matrix comparison method as described here, and the matrices turned out to be correlated (Mantel's test $r=0.70, \: p=0.001$) (Fig. \ref{fig:validation-results}) as expected, thus validating that the distance matrix comparison method outlined above successfully identifies acoustic representations in the neural data.

\begin{equation} \label{eq:poisson}
P(x = k) = e^{-\mu}\dfrac{\mu^{k}}{k!}
\end{equation}
where \textit{\textmu} is the expected number of spikes per time step (mean spiking rate), \textit{P(x = k)} denotes the probability of firing \textit{k} spikes on any given time step.

\textbf{Mantel Test} To test for correlation between a pair of distance matrices, we use the Mantel Test. It is a permutation test specifically designed to test the correlation between distance matrices \cite{Mantel_1967}. We cannot use a standard correlation coefficient as distance values are not independent of each other.

\subsection{Classification using acoustic vs. semantic labels}
In this analysis, we wanted to compare the amount of semantic and acoustic information available in A1 and A2. Here, we represent the neural population response to a given stimulus as a \textit{T} x \textit{N} (where \textit{T} denotes timesteps and \textit{N} denotes neurons) matrix that describes the time-varying firing rates of neurons comprising the population. We used dynamic representations of neural activity here as it has been shown to contain significantly more discriminatory information \cite{Robotka_Thomas_Yu_Wood_Elie_Gahr_Theunissen_2023}. 

The stimuli were labeled in two ways. The first approach (acoustic labels) involved hierarchically clustering the stimulus waveforms based on their mel-frequency cepstral coefficients (MFCCs) and using the cluster assignments as labels \cite{Robotka_Thomas_Yu_Wood_Elie_Gahr_Theunissen_2023}. The second approach (semantic labels) used the provided human-annotated labels that took into account the behavioral context in which the calls were emitted. 

Classification performance was measured for A1 and A2 separately for both the label assignments. A time-series Support Vector Machine (SVM) classifier with a linear kernel was used to classify neural representations. Linear Discriminant Analysis (LDA) was used to reduce the dimensionality of the population response matrices before being passed on to the SVM classifier. Classification performance was defined as accuracy score (percentage correct classifications, or PCC) and was estimated using 3-fold cross validation.

\subsection{Dimensionality of representation spaces}
We hypothesized that the representation space of A2 should be higher-dimensional as compared to A1, in order to support greater separation of semantic categories. Here, we use spike-count vectors to represent the neural population response. We define the embedding dimensionality of the representation space of a neural population as the number of linear dimensions required to capture 80\% variance of the \textit{S} x \textit{N} matrix, where \textit{S} denotes stimuli and \textit{N} denotes neurons \cite{Lehky_Kiani_Esteky_Tanaka_2014, Humphries_2021}. The entries of the matrix contain spike-counts of neuron \textit{n} in response to stimulus \textit{s}. Principal Component Analysis (PCA) was used to estimate the embedding dimensionality.

\subsection{Separation of semantic categories in A1 vs. A2}
We hypothesized that semantic categories will be more separated in A2 as compared to A1. Here, we use spike-count vectors to represent the neural population response. For a given semantic category, we define separation as the difference in means of across-categories distances and within-category distances, normalized by the standard deviation of the within-category distances (Eq. \ref{eq:separation}) \cite{Fisher_1936}. Here, we used the cosine distance so that the distance between two population response vectors is invariant to the magnitude of the responses (i.e., the number of spikes emitted).
\begin{equation} \label{eq:separation}
s_{category} = \dfrac{\mu_{across\_categories} - \mu_{within\_category}}{\sigma_{within\_category}}
\end{equation}

\section{Results}
\label{sec:results}

\subsection{Distance matrix comparison}
\begin{figure}[t]
  \centering
  \includegraphics[width=\linewidth]{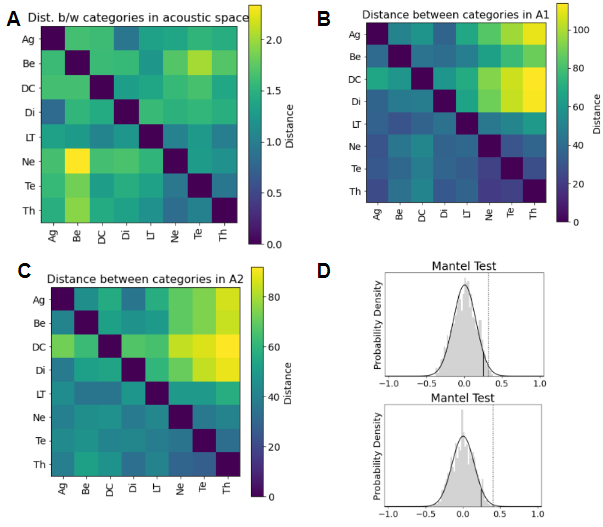}
  \caption{The distances between semantic categories in the (A) acoustic feature space is correlated with the corresponding distances in either (B) the A1 representation space or (C) the A2 representation space. (D) Observed correlation values visualized on the histogram obtained from the permutations of the Mantel Test. (Top) Correlation between (A) and (B). (Bottom) Correlation between (A) and (C).}
  \label{fig:distance-matrices}
\end{figure}

We found that the distances between semantic categories in the acoustic feature space are correlated with the corresponding distances in either the A1 or the A2 representation space (Mantel's test $r(A1)=0.319, \: p(A1) = 0.008, r(A2)=0.401, \: p(A2)=0.001$) (Fig. \ref{fig:distance-matrices}). This indicates that acoustic features are encoded in the responses of both A1 and A2.

\subsection{Classification using acoustic vs. semantic labels}
\begin{figure}[ht]
  \centering
  \includegraphics[width=\linewidth]{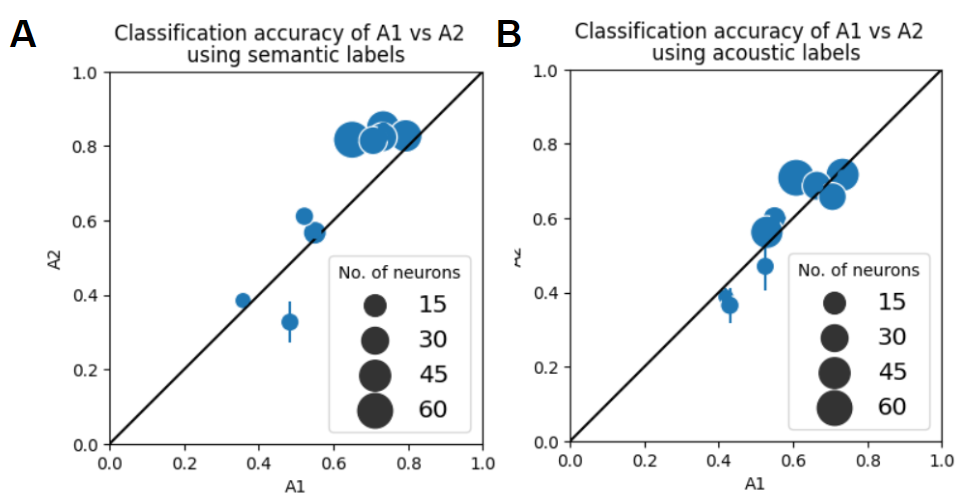}
  \caption{Acoustic and semantic classification in A1 vs A2. Each marker represents a given hemisphere from a given bird. Marker size indicates the neural population size. (A) Classification accuracy (PCC) for decoding (A) semantic category of calls and (B) acoustic category of calls using A1 and A2 activity. Markers on the diagonal indicate that for the given hemisphere, A1 and A2 are equally informative.}
  \label{fig:acoustic-vs-semantic}
\end{figure}

We compute semantic decoding accuracy (as given by PCC) for A1 and A2 in all hemispheres. We find that consistently across hemispheres, A2 activity is more informative of the semantic category of vocalizations ($z=1.717, \: p=0.043$, Wilcoxon Signed-Rank Test) (Fig. \ref{fig:acoustic-vs-semantic}A). Notice that larger population sizes correspond to higher gains for A2. Ruling out the possibility that the observed differences are a result of better acoustic representations in A2, we report that A1 and A2 are equally informative of the acoustic category of the stimuli ($z=0.113, \: p=0.910$, Wilcoxon Signed-Rank Test) (Fig. \ref{fig:acoustic-vs-semantic}B).

\subsection{Dimensionality of representation spaces}
We calculated the embedding dimensionality for A1 and A2 in each hemisphere. We find that A2 consistently uses a higher-dimensional representation space as compared to A1 ($z=1.787, \: p=0.037$, Wilcoxon Signed-Rank Test) (Fig. \ref{fig:separation}A). Higher-dimensional spaces have the potential to achieve greater separation of categories \cite{Beyeler_Rounds_Carlson_Dutt_Krichmar_2019}.

\subsection{Separation of semantic categories in A1 and A2}
\begin{figure}[ht]
  \centering
  \includegraphics[width=\linewidth]{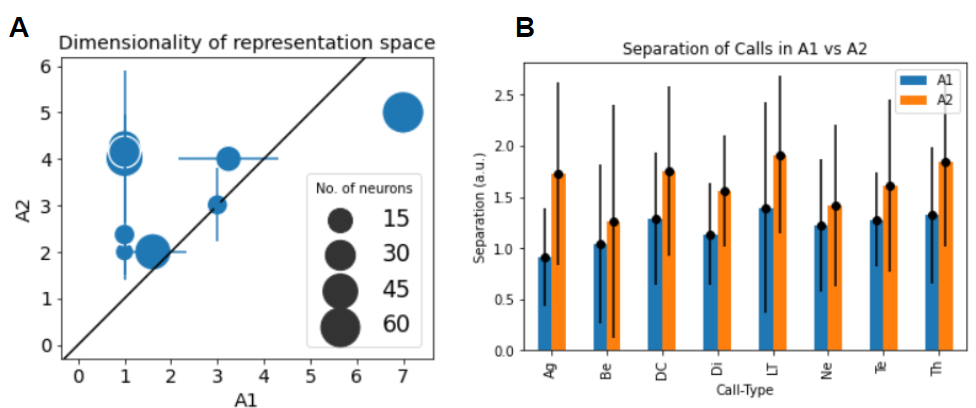}
  \caption{(A) Embedding dimensionality of representation spaces of A1 and A2, given by number of linear dimensions required to capture 80\% variance. Each marker represents one hemisphere. (B) Separation of semantic categories in the representation spaces of A1 vs A2.}
  \label{fig:separation}
\end{figure}
We calculated separation for each semantic category in A1 and A2. We found that semantic categories are more separated in A2 as compared to A1 ($z = 2.652, \: p = 0.004$, Wilcoxon Signed-Rank Test) (Fig. \ref{fig:separation}B). This further demonstrates that A2 is extracting semantic information from the input signal.

\section{Discussion}
\label{sec:discussion}

In this study, we hypothesized that more semantic information exists in A2 as compared to A1. We first demonstrated that acoustic features are encoded in the responses of both A1 and A2 by showing that the distance matrix of calls in the acoustic feature space is correlated with that in the representation spaces of A1 and A2. We then demonstrated that the meaning of vocalizations can be decoded with greater accuracy from the activity of A2. To rule out the possibility that the observed gain in accuracy is simply a result of a better acoustic representation, we established that both A1 and A2 are comparable in terms of their acoustic classification performance. Finally, we showed that A2 has a higher-dimensional representation space and that semantic categories are more separated in A2.

Note that we have used Euclidean distance to describe distances between vectors for the distance matrix comparison and cosine distance for the separation comparison. This is because in the distance matrix comparison, Hausdorff distance was used to describe the distances between semantic categories, and it can only be used with a true distance measure (i.e., one that satisfies the triangle inequality). However, there are no such restrictions in the separation analysis. Therefore, we opted for cosine distance as it is invariant to the magnitude of responses. 

Previously reported results \cite{Robotka_Thomas_Yu_Wood_Elie_Gahr_Theunissen_2023} are inconsistent with what we already know about A1 and A2. From lesion studies, we know that while unilateral lesions of the A1 lead to slight hearing loss, unilateral lesions of the A2 can lead to a loss in auditory language comprehension ability \cite{Mendoza_2011}. Therefore, we attribute the findings of Robotka et al (2023) \cite{Robotka_Thomas_Yu_Wood_Elie_Gahr_Theunissen_2023} to the small neural ensemble sizes used in their analysis. In our own study, we found that the difference of A2 accuracy scores and A1 accuracy scores when decoding semantic labels is significantly correlated with the neural population size (Pearson's $r = 0.637, \: p = 0.032$). Therefore, we believe that our work, using much larger populations, paints a more accurate picture of the roles of A1 and A2.

In their work, Robotka et al (2023) \cite{Robotka_Thomas_Yu_Wood_Elie_Gahr_Theunissen_2023} conclude that their findings are consistent with two models of information processing in the auditory system: first, that downstream regions combine information from both regions (which they deem unlikely), and second, that either of the two regions could be the source of downstream information flow. Based on our findings, we propose that A2 is the likely source of downstream information flow in tasks that require semantic understanding.

\vfill\pagebreak
\bibliographystyle{IEEEbib}
\bibliography{refs}

\begin{thebibliography}{10}

\bibitem{Robotka_Thomas_Yu_Wood_Elie_Gahr_Theunissen_2023}
H.~Robotka, L.~Thomas, K.~Yu, W.~Wood, J.E. Elie, M.~Gahr, and F.E. Theunissen,
\newblock ``Sparse ensemble neural code for a complete vocal repertoire,''
\newblock {\em Cell Reports}, vol. 42, no. 2, pp. 112034, Jan 2023.

\bibitem{Elie_Theunissen_2015}
Julie~E. Elie and Frédéric~E. Theunissen,
\newblock ``Meaning in the avian auditory cortex: neural representation of communication calls,''
\newblock {\em European Journal of Neuroscience}, vol. 41, no. 5, pp. 546–567, Mar 2015.

\bibitem{Elie_Theunissen_2019}
Julie~E. Elie and Frédéric~E. Theunissen,
\newblock ``Invariant neural responses for sensory categories revealed by the time-varying information for communication calls,''
\newblock {\em PLOS Computational Biology}, vol. 15, no. 9, pp. e1006698, Sep 2019.

\bibitem{Meliza_Margoliash_2012}
C.~D. Meliza and D.~Margoliash,
\newblock ``Emergence of selectivity and tolerance in the avian auditory cortex,''
\newblock {\em Journal of Neuroscience}, vol. 32, no. 43, pp. 15158–15168, Oct 2012.

\bibitem{Elie_Theunissen_2016}
Julie~E. Elie and Frédéric~E. Theunissen,
\newblock ``The vocal repertoire of the domesticated zebra finch: a data driven approach to decipher the information-bearing acoustic features of communication signals,''
\newblock {\em Animal cognition}, vol. 19, no. 2, pp. 285–315, Mar 2016.

\bibitem{Tsunada_Cohen_2014}
Joji Tsunada and Yale~E. Cohen,
\newblock ``Neural mechanisms of auditory categorization: from across brain areas to within local microcircuits,''
\newblock {\em Frontiers in Neuroscience}, vol. 8, 2014.

\bibitem{Elie_Theunissen_Dataset}
Julie~E. Elie and Frédéric~E. Theunissen,
\newblock ``Simultaneous extracellular recordings of avian auditory neurons in zebra finches presented with all the repertoire of vocalizations used by this species for vocal communication,''
\newblock Tech. {R}ep., CRCNS.org, 2019,
\newblock doi:http://dx.doi.org/10.6080/K00C4T06.

\bibitem{Kriegeskorte_Mur_Bandettini_2008}
Nikolaus Kriegeskorte, Marieke Mur, and Peter Bandettini,
\newblock ``Representational similarity analysis - connecting the branches of systems neuroscience,''
\newblock {\em Frontiers in Systems Neuroscience}, vol. 2, 2008.

\bibitem{Taha_Hanbury_2015}
Abdel~Aziz Taha and Allan Hanbury,
\newblock ``An efficient algorithm for calculating the exact hausdorff distance,''
\newblock {\em IEEE Transactions on Pattern Analysis and Machine Intelligence}, vol. 37, no. 11, pp. 2153–2163, Nov 2015.

\bibitem{Mantel_1967}
N~Mantel,
\newblock ``The detection of disease clustering and a generalized regression approach,''
\newblock {\em Cancer research}, vol. 27, no. 2, pp. 209–220, Feb 1967.

\bibitem{Lehky_Kiani_Esteky_Tanaka_2014}
Sidney~R. Lehky, Roozbeh Kiani, Hossein Esteky, and Keiji Tanaka,
\newblock ``Dimensionality of object representations in monkey inferotemporal cortex,''
\newblock {\em Neural Computation}, vol. 26, no. 10, pp. 2135–2162, Oct 2014.

\bibitem{Humphries_2021}
Mark~D. Humphries,
\newblock ``Strong and weak principles of neural dimension reduction,''
\newblock {\em Neurons, Behavior, Data Analysis and Theory (NBDT)}, vol. 5, no. 2, Jun 2021.

\bibitem{Fisher_1936}
R.~A. Fisher,
\newblock ``The use of multiple measurements in taxonomic problems,''
\newblock {\em Annals of Eugenics}, vol. 7, no. 2, pp. 179–188, 1936.

\bibitem{Beyeler_Rounds_Carlson_Dutt_Krichmar_2019}
Michael Beyeler, Emily~L. Rounds, Kristofor~D. Carlson, Nikil Dutt, and Jeffrey~L. Krichmar,
\newblock ``Neural correlates of sparse coding and dimensionality reduction,''
\newblock {\em PLOS Computational Biology}, vol. 15, no. 6, pp. e1006908, Jun 2019.

\bibitem{Mendoza_2011}
John~E. Mendoza,
\newblock ``Auditory cortex,''
\newblock in {\em Encyclopedia of Clinical Neuropsychology}, Jeffrey~S. Kreutzer, John DeLuca, and Bruce Caplan, Eds., p. 301–301. Springer New York, New York, NY, 2011.

\end{thebibliography}

\end{document}